\documentclass[10pt,a4paper]{article}
\usepackage{epsfig}
\usepackage{changes}
\usepackage{latexsym}
\columnwidth\textwidth

\newcommand{\swb}{\small\bf\tt }
\newcommand{\swn}{\small\tt }
\newcommand\abk[1]{{\rm #1}}
\begin{document}

\begin{center}
\LARGE
{\bf Neutrino Astronomy: A New Window to the Universe} \\~\\
\end{center}
\Large
\begin{center}
{\bf Claus Grupen}

\normalsize
{\it University of Siegen}
\end{center}
\vspace{10mm}
\setlength {\textwidth}{100mm}
\large
\begin{center}
{\bf ABSTRACT} \\
\begin{minipage}[t]{115mm}
\small
Neutrino astronomy offers the possibility to look into the interior of
astrophysical objects. This advantage over the observation in classical
astronomy in various regions of the electromagnetic spectrum goes along with a
difficult detection of neutrinos. Pioneering experiments already have seen the
sun and the supernova 1987A in the light of neutrinos. This offers the prospects
to be able to look in the future into compact astrophysical objects which may be
the sources of cosmic radiation. \\~\\
\end{minipage}
\end{center}

\begin{figure} [hbt]
\begin{center}
{\epsfig{figure=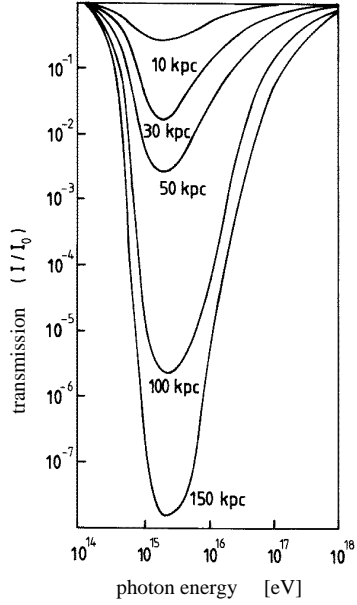,width=0.5
\textwidth}}
\caption[]
 {\tt Fractional absorption of high energy $\gamma$-rays by the 2.7 K 
 black\-body radiation for different cosmic distances \cite{ramana, cawley}.}
\label{gamma}
\end{center}
\end{figure}

\enlargethispage{30mm}
\noindent
\large
{\bf INTRODUCTION} \\
\normalsize
The disadvantage with "classical astronomies", such as the observation in the
radio, infrared, optical, ultraviolet, X-ray or $\gamma$-ray regime is related
to the fact that electromagnetic radiation is rapidly absorbed in matter and
therefore only the surfaces of astronomical objects are visible. In addition,
energetic $\gamma$-rays from distant sources are absorbed by $\gamma$-$\gamma$
interactions with blackbody photons through the process
\begin{displaymath}
\gamma + \gamma \rightarrow \abk{e}^{+} + \abk{e}^{-}.
\end{displaymath}
The threshold energy for this process is around $E_{\gamma}=10^{14}\,$eV.
Energetic photons from the Large Magellanic Cloud (LMC, distance $55\,$kpc) are
already substantially absorbed by this process (Figure \ref{gamma}).

Charged primary cosmic rays could in principle also be used in astroparticle
physics. The directional information, however, is only preserved for very
energetic protons or nuclei, because at lower energies irregular magnetic fields
randomize their arrival direction. On the other hand, energetic protons also
interact with blackbody or starlight photons thereby degrading their energy. For
protons of energies in excess of $5 \cdot 10^{19}\,$eV the universe is no longer
transparent.

The requirements for a good astronomy are:
\begin{enumerate}
\item particles (or radiation) must not be affected by regular or irregular
magnetic fields.
\item particles must arrive at Earth. This excludes unstable particles, such as
neutrons, unless their energy is extremely high.
\vspace{10mm}
\large
\begin{center}
\hspace{-10mm}Conf.~Proc.~XV Cracow Summer School on Cosmology, July 15-19, 
1996, Lodz, Poland
\end{center}
\normalsize
\item particle and antiparticle should be distinguishable. This excludes
photons, because $\gamma = \bar{\gamma}$.
\item the particles must be penetrating to allow to look into the interior of
stellar or galactic objects.
\item the particles should not be absorbed e.g.~by blackbody photons.
\end{enumerate}

Neutrinos fulfil all these requirements.

\pagestyle{plain}
\noindent
\\
\\
\large
{\bf NEUTRINO BASICS} \newline
\normalsize
In the Standard Model of elektroweak interactions there are three families of
quarks and leptons:
\begin{center}
\vspace{-3mm}
\begin{displaymath}
\abk{quarks} \hspace{5mm} {\abk{u} \choose \abk{d}} \hspace{7.5mm} {\abk{c} 
\choose \abk{s}}
\hspace{7.5mm} {\abk{t} \choose \abk{b}}
\end{displaymath} 
\end{center}

\begin{center}
\begin{displaymath}
\hspace{2mm}\abk{leptons} \hspace{4mm} {\nu_{\abk{e}} \choose \abk{e}^{-}} 
\hspace{5mm} {\nu_{\mu} \choose \mu^{-}} \hspace{5mm} {\nu_{\tau} \choose 
\tau^{-}}
\end{displaymath}
\end{center}

The three neutrinos could have zero mass. Experimentally one can provide only
mass limits \cite{pdg}:

\vspace{3mm}
\begin{math}
\hspace{10mm}m_{\nu_{\abk{e}}} \le 4.5\, \abk{eV} \hspace{12.5mm} \abk{from~} 
^{3}\abk{H-decay}
\end{math}

\begin{math}
\hspace{10mm}m_{\nu_{\mu}} \le 270\, \abk{keV} \hspace{9.5mm} \abk{from~}
\pi^{+} \rightarrow \mu^{+} + \nu_{\mu}\abk{~~decay}
\end{math}

\begin{math}
\hspace{10mm}m_{\nu_{\tau}} \le 24\, \abk{MeV} \hspace{10mm} \abk{from~}
\tau^{+} \rightarrow
3\, \pi^{+} + 2\, \pi^{-} + \bar{\nu}_{\tau}\abk{~~decay}
\end{math}
\\
\\
\large
{\bf ATMOSPHERIC NEUTRINOS} \newline
\normalsize
For neutrino astronomy atmospheric neutrinos are an annoying background.
However, for the study of interactions and in search for possible neutrino
oscillations they may be interesting in their own right. A na\"{\i}ve 
expectation 
for the $\nu_{\mu}/\nu_{\abk{e}}$ ratio for atmospheric neutrinos can be 
derived from their main sources

\begin{displaymath}
\pi^{+} \rightarrow \mu^{+} + \nu_{\mu}, \hspace{3mm} \pi^{-} \rightarrow
\mu^{-} + \bar{\nu}_{\mu}
\end{displaymath}
\vspace{-5mm}
\begin{displaymath}
\mu^{+} \rightarrow \abk{e}^{+} + \bar{\nu}_{\mu} + \nu_{\abk{e}}, \hspace{3mm}
\mu^{-} \rightarrow \abk{e}^{-} + {\nu_{\mu}} + \bar{\nu}_{\abk{e}}.
\end{displaymath}

One would expect a ratio of 

\begin{displaymath}
\frac{N \left(\nu_{\mu}, \bar{\nu}_{\mu} \right)} {N \left(\nu_{\abk{e}},
\bar{\nu}_{\abk{e}} \right)} = 2.
\end{displaymath}

\begin{figure} [ht]
\begin{center}
{\epsfig{figure=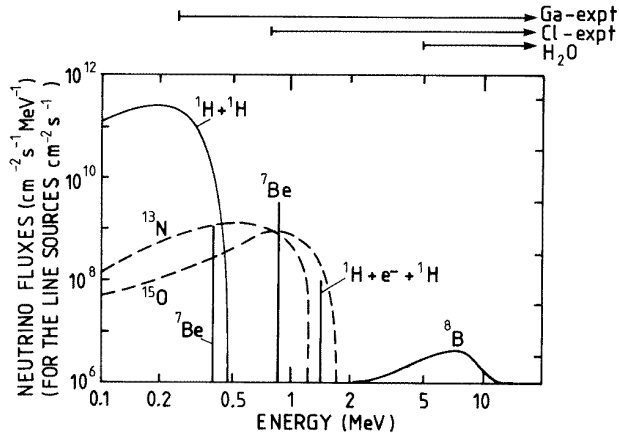,height=63mm}}
\caption[]
 {\tt Theoretical differential energy spectra of electron neutrinos from nuclear
 reactions in the interior of the sun \cite{bahcall}. Also indicated are the
 threshold energies for neutrino detection in the Chlorine, Gallium and
 Water-Cherenkov experiments}
\label{spectra}
\end{center}
\end{figure}

Some experiments find a deficit of muon-type neutrinos \cite{kajita} and some do
not \cite{daum}. In view of the difficult detection of low energy muon neutrinos
one should be rather reluctant to propose new physics to explain a possible
discrepancy.
\\
\\
\large
{\bf SOLAR NEUTRINOS} \newline
\normalsize
The majority of neutrinos in the sun is produced in the proton-proton fusion
reaction p+p $\rightarrow$ d + e$^{+}$ + $\nu_{\abk{e}}$ ("pp-neutrinos", 86\%).
About 14\% originate from the electron capture process $^{7}$Be + e$^{-}$
$\rightarrow$ $^{7}$Li + $\nu_{\abk{e}}$. A very small fraction (0.02\%) of
energetic neutrinos comes from the beta-decay of $^{8}$B: $^{8}$B $\rightarrow$ 
$^{8}$Be$^{\ast}$ + e$^{+}$ + $\nu_{\abk{e}}$. The total flux of neutrinos from
the sun is about $7 \cdot 10^{10}\,$cm$^{-2}$s$^{-1}$. The various sources
contributing to solar neutrino spectrum are shown in Figure \ref{spectra}.

Also indicated in this diagram are the detection threshold energies in the
various experiments looking for solar neutrinos.
These experiments are \cite{bahcall, gallex, koshiba}
\begin{enumerate}
\item Davis experiment: \hspace{8mm}$\nu_{\abk{e}} + ^{37}$Cl $\rightarrow$ 
e$^{-} + ^{37}$Ar \hspace{4mm} $\left(E_{\nu} \ge 810\,\abk{keV} \right)$
\item GALLEX and SAGE: \hspace{1.5mm} $\nu_{\abk{e}} + ^{71}$Ga $\rightarrow$ 
e$^{-} + ^{71}$Ge \hspace{3mm} $\left(E_{\nu} \ge 233\,\abk{keV} \right)$
\item Kamiokande: \hspace{14.5mm} $\nu_{\abk{e}}$ + e$^{-}$ $\rightarrow$ 
$\nu_{\abk{e}}$ + e$^{-}$ \hspace{5.5mm} $\left(E_{\nu} \ge 5\,\abk{MeV} 
\right)$
\end{enumerate}

In the first two cases of radiochemical experiments the minute number of 
produced $^{37}$Ar or $^{71}$Ge atoms has to be extracted in complicated
chemical procedures and carefully counted. These integrating experiments 
provide no directional
information, in  contrast to the Kamiokande experiment, where the measured
electron direction can be related to the position of the sun. The results of the
different experiments are shown in Table 1.
\clearpage

\begin{table}[hb]
\begin{center}
\begin{tabular}{|c|c|c|c|}
\hline
 {\swb Experiment} &  {\swb Results} & \multicolumn{2}{c|}{\swb Prediction}\\
 \raisebox{1.5ex}[-1.5ex]  &  & {\swb Bahcall } & {\swb Turck-Chi\`eze}\\
\hline
\hline
{\swn Davis et al.} & {\swn 2.55 $\pm$ 0.25} & {\swn 8.0 $\pm$ 1.0} & 
{\swn 6.4 $\pm$ 1.4}\\
\hline
{\swn Kamiokande} & {(\swn 0.51 $\pm$ 0.07) $\times$ predicted} & 
{\swn 1.0 $\pm$ 0.14} & {\swn 0.65 $\pm$ 0.07} \\
\hline
{\swn SAGE} & {\swn 73 $\pm$ 18 $\pm$ 7} & {\swn 131.5 $\pm$ 7} & {\swn 122.5 
$\pm$ 7} \\
\hline
{\swn GALLEX} & {\swn 76.7 $\pm$ 8.4 $\pm$ 4.9} & {\swn 131.5 $\pm$ 7} & {\swn 
122.5 $\pm$ 7} \\
\hline
\end{tabular}
\caption[]{\swn Observed and predicted rates of neutrino fluxes 
\hspace{2mm}\cite{gallex, moessbauer}.
The Kamiokande result and the Turck-Chi\`eze prediction are normalized to the 
Bahcall value.}
\label{komposition}
\end{center}
\end{table}

For Davis et al., SAGE and GALLEX the rates are given in {\em solar neutrino 
units} (1 SNU = one capture per second per picobarn = neutrino flux $\times$
cross-section $\left[10^{-36}\, \abk{s}^{-1} \right] )$

It appears that the "pp-neutrinos" are seen by SAGE and GALLEX and that there
is a problem with the $^{7}$Be and $^{8}$B neutrinos. Many ideas have been put
forward to solve the problem of missing solar neutrinos:
\begin{enumerate}
\item is the standard solar model correct? The flux of $^{8}$B-neutrinos varies
with the central temperature of the sun like $T^{18}$. A small decrease of this
temperature  could solve the $^{8}$B-neutrino deficit. Turbulence of the solar
material or observations in helioseismology indicate that the standard model may
have to be modified. A more exotic explanation could be provided by WIMPs which
could have been trapped by the sun thereby lowering its core temperature
\cite{trefil}.
\item the apparent deficit of $^{7}$Be-neutrinos could be understood by an
overestimated cross-section for $^{7}$Be production in the sun at low energies.
\item a more exotic but very popular interpretation of the neutrino deficit (if
it is real) could be provided by neutrino oscillations \cite{bahcall,langacker,
morrison}. If the neutrinos were not
massless they could oscillate from one neutrino flavour to the other. Since the
solar neutrino detectors are only sensitive to $\nu_{\abk{e}}$, a maximal mixing
could lead to equal amounts of $\nu_{\abk{e}}$, $\nu_{\mu}$ and $\nu_{\tau}$ at
Earth and hence to a detection rate of only one third of the original neutrino
flux from the interior of the sun. There are two varieties of neutrino mixing:
one can have either neutrino-oscillation in vacuum or matter enhanced
oscillations. The matter oscillations depend on the electron density in the sun.
Somewhere inside the sun the electron density could be just right to induce
resonant neutrino matter oscillations.
\item if neutrinos had a finite mass they could also have a magnetic moment. If
their spin is flipped from the site of production to the detector on Earth they 
will not be seen because the neutrino detectors are insensitive to neutrinos 
with wrong helicity.
\end{enumerate}

There are also more exotic explanations for the apparent solar neutrino 
deficit. I would prefer to check the points (1.) and (2.) before new phenomena
are advocated.

\pagestyle{plain}
\noindent
\\
\\
\large
{\bf SUPERNOVA NEUTRINOS} \newline
\normalsize
The star Sanduleak exploded in the Large Magellanic Cloud in 1987. In supernova
explosions vast numbers of neutrinos are emitted. There are two sources of
neutrinos: the first comes from the deleptonization process p + e$^{-}$
$\rightarrow$ $\nu_{\abk{e}}$ + n when the neutron star is formed, the second
source are thermal neutrinos which are produced at temperatures of around 
$10\,$MeV through the chain
\vspace{-1mm}
\begin{displaymath}
\gamma + \abk{N} \rightarrow \abk{e}^{+} + \abk{e}^{-} + 
\tilde{\abk{N}}
\end{displaymath}
\vspace{-5mm}
\begin{displaymath}
\abk{e}^{+} + \abk{e}^{-} \rightarrow \abk{Z} \rightarrow
\nu_{\abk{x}} + \bar{\nu}_{\abk{x}} \hspace{4mm} \abk{x} = \abk{e}, \mu, \tau
\hspace{2mm}.
\end{displaymath}

Two experiments (Kamiokande and IMB) have detected neutrinos from SN 1987A. The 
IMB-experiment saw 8 events (threshold $\ge 19\,$MeV) while Kamiokande saw 12
events (threshold $\ge 5\,$MeV) \cite{koshiba}. Mainly $\bar{\nu}_{\abk{e}}$'s
were detected via the charged current process $\bar{\nu}_{\abk{e}} + \abk{p}
\rightarrow \abk{n} + \abk{e}^{+}$.
The total energy emitted in the form of neutrinos was estimated to be
$E_{\abk{total}}=\left(6 \pm 2 \right) \cdot 10^{46}\,$J corresponding to a 
total\linebreak
\clearpage
\noindent
neutrino flux of $\sim 10^{58}$ emitted over a time of $\sim 10$ seconds.
From the fact that the supernova neutrinos arrived at Earth a limit for the
neutrino lifetime could be derived, and the observed time dispersion at Earth
was used to infer an upper limit on the electron neutrino mass of
$m_{\nu_{\abk{e}}} \le 10\,$eV.
\noindent
\\
\\
\large
{\bf GALACTIC AND EXTRAGALACTIC NEUTRINOS} \newline
\normalsize
The most popular acceleration and production mechanism of energetic galactic or
extragalactic neutrinos is from binaries consisting of a "target" star which is
orbited by a "production" pulsar. The pulsar accelerates protons which interact
in the stellar atmosphere of the companion star according to 

\begin{center}
\begin{math}
\abk{p} + \abk{nucleus} \rightarrow \pi^{+}\quad +\quad \pi^{-}\quad +\quad 
\pi^{0}\quad +\quad \abk{X}
\end{math}
\end{center}
\vspace{-3mm}
\hspace{49.5mm}
\put(144,0){\line(0,1){12}}
\put(144,0){\vector(1,0){12}}

\vspace{-3.9mm}
\begin{math}
\hspace{51mm}\mu^{+}\ \nu_{\mu}
\end{math}

\vspace{-5.8mm}
\hspace{60mm}
\put(175,0){\line(0,1){25}}
\put(175,0){\vector(1,0){11}}

\vspace{-3.9mm}
\begin{math}
\hspace{66.5mm}\mu^{-}\ \bar{\nu}_{\mu}
\end{math}

\vspace{-10mm}
\hspace{76mm}
\put(220,0){\line(0,1){35}}
\put(220,0){\vector(1,0){12}}

\vspace{-3.9mm}
\begin{math}
\hspace{82.5mm}\gamma\ \gamma 
\end{math}

\vspace{3mm}
\noindent
providing equal amounts of neutrinos and $\gamma$-rays. The photon yield,
however, strongly depends on the pulsar phase and the interplay of the local
density and the column density of the stellar atmosphere \cite{stenger}. 
But also more unorthodox production mechanisms involving cosmic strings are 
proposed \cite{sige}.

\begin{figure} [hbt]
\begin{center}
{\epsfig{figure=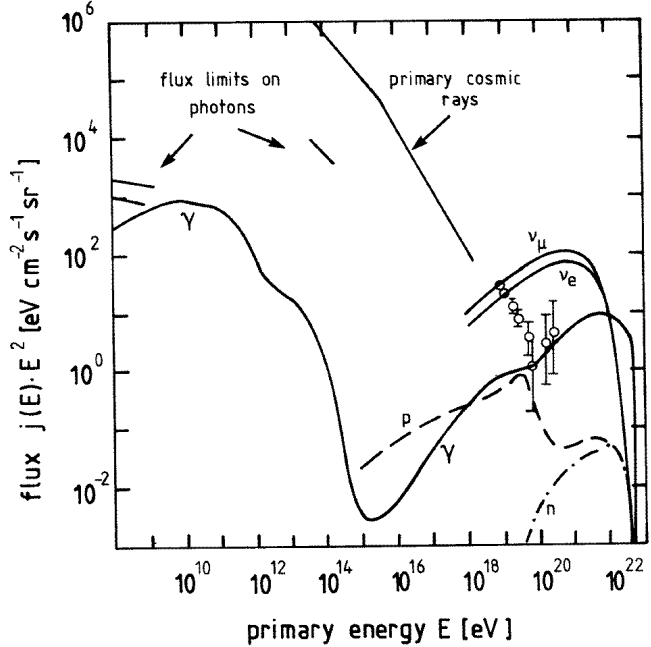,height=0.92
\textwidth}}
\caption[]
 {\tt Predictions for the differential fluxes of $\gamma$-rays, neutrinos,
 protons and neutrons from the collapse or annihilation of topological defects,
 such as cosmic strings via the decay of GUT X-particles along with experimental
 data on charged cosmic ray rates and upper limits on $\gamma$-ray fluxes
(\cite{sige} and references therein).}
\label{drei}
\end{center}
\end{figure}

In these models particles are created at ultrahigh energies ($10^{24}\,$eV) by
the decay of massive X-particles associated with new fundamental unified
interactions near the grand unification (GUT) scale. Such gauge theories predict
phase transitions in the early universe which are expected to create topological
defects, such as e.g.~cosmic strings. These cosmic strings could possibly
release X-particles due to collapse or annihilation processes \cite{sige}.
In GUT-theories the X-particles are predicted to decay into jets of hadrons,
which would eventually provide copious numbers of $\gamma$-rays and neutrinos.
Figure \ref{drei} shows expectations for the differential fluxes of 
$\gamma$-rays,
neutrinos, protons and neutrons based on the cosmic string origin \cite{sige}.
The fluxes have been estimated for spatial uniform injection; i.e.~the
particles were propagated through extragalactic space, and the fluxes were
normalized to the observed particle rate at $10^{20}\,$eV. Figure \ref{drei}
also shows
experimental data from AGASA and the Fly's Eye experiment (dots with error
bars), piecewise power law fits to the observed charged cosmic ray rate and
experimental upper limits on the  $\gamma$-ray flux (\cite{sige} and references
therein). 

For neutrino detectors under construction (DUMAND, AMANDA, NESTOR) the minimum
detectable flux for neutrinos in the TeV region is of the order of $10^{-10}\,
$cm$^{-2}$s$^{-1}$ \cite{stenger2, halzen, resvanis}. This limit could possibly
be somewhat pessimistic if one assumes that the neutrino nucleon cross-section
rises substantially at large energies due to the abundance of low energy 
partons inside the nucleon \cite{ralston}. Candidate sources which might fulfil
the minimum flux requirement are VELA X1, CRAB, Cyg X3, Cen A, Markarian 421 
and the quasar
3C273. Also all binary pulsars, supernova shells, active galactic nuclei and the
galactic center are possible candidates. Up to now, however, no point source
emitting high energy neutrinos has been detected.
\noindent
\\
\\
\large
{\bf CONCLUSION} \newline
\normalsize
Atmospheric neutrinos seem to be well under control. The discrepancy between the
predicted and actually measured number of solar neutrinos  can probably be
understood by minor modifications to the standard solar model and by use of
improved measurements of nuclear processes relevant for neutrino production in
the solar core. The supernova neutrinos are in exellent shape. Energetic
galactic or extragalactic neutrinos have not been seen yet. Ongoing accelerator
experiments will answer the question whether neutrino oscillations are a
reality.
\noindent
\\
\\
\large
{\bf ACKNOWLEDGEMENTS} \newline
\normalsize
I am grateful for the hospitality and support provided by the summer school
organizers. My special thanks go to Janusz Kempa, Jerzy Wdowczyk and Wieslaw
Tkaczyk. I thank also Volker Schreiber and Detlev Maier for their help in 
preparing the written version of my talk.

\end{document}